\documentclass[]{lni}
\usepackage{graphicx}
\usepackage[nolist]{acronym}
\usepackage{tikz}
\begin{document}

%%% Mehrere Autoren werden durch \and voneinander getrennt.
%%% Die Fußnote enthält die Adresse sowie eine E-Mail-Adresse.
%%% Das optionale Argument (sofern angegeben) wird für die Kopfzeile verwendet.

\title[PrivacyScore]{PrivacyScore: Analyse von Webseiten auf Sicherheits- und Privatheitsprobleme – Konzept und rechtliche Zulässigkeit\textsuperscript{*}}
\subtitle{Dieses Papier ist eine erweiterte Fassung des Beitrags, der beim Workshop „Recht und Technik“ auf der INFORMATIK 2017 vorgestellt wird.} % if needed
\author[Max Maass \and Anne Laubach \and Dominik Herrmann] 
{Max Maass\footnote{Technische Universität Darmstadt, Secure Mobile Networking Lab, Mornewegstr. 32, 64293 Darmstadt, Deutschland,
\email{mmaass@seemoo.tu-darmstadt.de}} \and
Anne Laubach\footnote{Universität Kassel, Projektgruppe verfassungsverträgliche Technikgestaltung (provet), Pfannkuchstr. 1, 34109 Kassel, Deutschland, \email{a.laubach@uni-kassel.de}} \and
Dominik Herrmann\footnote{Universität Hamburg, Sicherheit verteilter Systeme, Vogt-Kölln-Str. 30, 22527 Hamburg,
Deutschland, \email{herrmann@informatik.uni-hamburg.de}}}
%\startpage{1} % Beginn der Seitenzählung für diesen Beitrag / Start page
%\editor{Herausgeber et al.} % Names of Editors
%\booktitle{INFORMATIK 2017 Workshop Recht und Technik: Datenschutz im Diskurs} % Name of book title
%\year{2017}

\maketitle
%TODO Nicht hübsch.
{
	\renewcommand{\thefootnote}{\fnsymbol{footnote}}
	\footnotetext[1]{Dieser Artikel ist in Teilen eine Übersetzung eines Papiers, das im Juni 2017 auf dem ENISA Annual Privacy Forum in Wien vorgestellt wurde \cite{Maass2017}. Neu hinzugekommen sind die juristischen Betrachtungen in Abschnitt~\ref{sec:legal}. Die offizielle Version dieses Artikels wird auf link.springer.com erscheinen, ein Link wird eingefügt, solbald dieser verfügbar ist.}
}

\begin{acronym}
	\acro{CDN}{Content-Distribution-Network}
	\acro{CMS}{Content-Management-System}
	\acrodefplural{CMS}{Content-Management-Systeme}
	\acro{DKIM}{DomainKey Identified Mail}
	\acro{DMARC}{Domain-based Message Authentication, Reporting and Conformance}
	\acro{DNS}{Domain-Name-System}
	\acro{DoS}{Denial-of-Service}
	\acro{DPA}{Data-Protection-Authority}
	\acrodefplural{DPA}{Data-Protection-Authoritys}
	\acro{DPO}{Data-Protection-Officer}
	\acro{HSTS}{HTTP Strict Transport Security}
	\acro{SPF}{Sender-Policy-Framework}
	\acro{TLS}{Transport-Layer-Security}
\end{acronym}

\begin{abstract}
PrivacyScore ist ein öffentliches Web-Portal, mit dem automatisiert überprüft werden kann, ob Webseiten gängige Mechanismen zum Schutz von Sicherheit und Privatheit korrekt implementieren. Im Gegensatz zu existierenden Diensten ermöglicht PrivacyScore, mehrere Webseiten in \emph{Benchmarks} miteinander zu vergleichen, die Ergebnisse differenziert und im Zeitverlauf zu analysieren sowie nutzerdefinierte Kriterien für die Auswertung zu definieren. PrivacyScore verbessert dadurch nicht nur die Transparenz für Endanwender, sondern erleichtert auch die Arbeit der Datenschutz"=Aufsichtsbehörden. In diesem Beitrag stellen wir das Konzept des Dienstes vor und wir erörtern, unter welchen Umständen das automatische Scannen und öffentliche „Anprangern“ von Schwächen aus rechtlicher Sicht zulässig ist.
\end{abstract}
%%% Beginn des Artikeltexts
\section{Einleitung}
Der sichere Betrieb einer Webseite ist eine Aufgabe, die viel technischen Sachverstand benötigt. 
Fahl \emph{et al.} haben gezeigt \cite{Fahl2014}, dass unsicher konfigurierte Webseiten oft eine Folge dieser Komplexität sind. 
Unsichere Webseiten gefährden nicht nur die Infrastruktur der Seitenbetreiber, sondern auch die Sicherheit und Privatheit der Besucherinnen.
Die Privatheit der Besucherinnen kann allerdings auch durch Entscheidungen des Seitenbetreibers beeinträchtigt werden, etwa durch die Verwendung von Analysediensten oder Werbenetzwerken. Seitenbetreiber nutzen hier oft kommerzielle Angebote (etwa \emph{Google Analytics}) anstatt einer lokalen Lösung, die die Privatheitsinteressen der Besucherinnen besser berücksichtigt (z.\,B. \emph{Piwik}).

Die Entscheidungen der Betreiber sind dabei für Besucherinnen oft intransparent, insbesondere bei komplexeren Diensten wie dem Schutz gegen Denial-of-Service-Angriffe. Für Besucherinnen ist hier teilweise nicht einmal die Verwendung eines solchen Dienstes erkennbar, obwohl er signifikanten Einfluss auf ihre Privatheit haben kann, was zuletzt die Sicherheitslücke bei dem Dienstleister \emph{Cloudflare} gezeigt hat.\footnote{\url{https://blog.cloudflare.com/incident-report-on-memory-leak-caused-by-cloudflare-parser-bug/}}

Es gibt zwar bereits einige kostenlose Webseiten"=Scanner, die die Konfiguration einer Webseite auf gängige Sicherheitslücken untersuchen; Aspekte der Privatheit werden von den existierenden Scannern jedoch weitgehend ignoriert. Des Weiteren bieten existierende Scanner keine Möglichkeit, die Ergebnisse mehrerer Webseiten direkt miteinander zu vergleichen.

Das PrivacyScore"=Projekt soll diese Lücke schließen. In Kooperation mit Datenschutz"=Aufsichtsbehörden und -Aktivisten entwickeln wir derzeit ein System, das die automatische Analyse und den Vergleich der Sicherheits- und Privatheitseigenschaften von Webseiten ermöglicht (\emph{Benchmark}). Wir haben drei Zielgruppen: \emph{Endnutzerinnen}, die sich über die Sicherheit und Privatheit bestimmter Webseiten informieren möchten, \emph{Wissenschaftlerinnen}, die große Mengen von Webseiten untersuchen wollen, und \emph{Aktivistinnen und Aufsichtsbehörden}, die die Erfüllung von gesetzlichen Datenschutz"=Mindeststandards durch Webseiten sicherstellen wollen.
Unsere Ideen und Realisierungsansätze haben wir erstmalig in \cite{Maass2017} vorgestellt. Im vorliegenden Papier betrachten wir zusätzlich die juristische Zulässigkeit der Erhebung der Daten und der Veröffentlichung von Benchmarks sowie die juristische Verwertbarkeit der Ergebnisse.

In Abschnitt~\ref{sec:relatedwork} stellen wir verwandte Arbeiten vor, bevor wir in Abschnitt~\ref{sec:overview} die PrivacyScore"=Plattform präsentieren. Die durchgeführten Tests werden in Abschnitt~\ref{sec:checks} beschrieben. In Abschnitt~\ref{sec:considerations} betrachten wir ethische Fragestellungen. Im Anschluss daran erörtern wir in Abschnitt~\ref{sec:legal} die Zulässigkeit aus juristischer Sicht. Abschnitt~\ref{sec:conclusion} enthält Schlussbemerkungen.

\section{Verwandte Arbeiten}
\label{sec:relatedwork}
Mehrere Dienste erlauben eine automatisierte Untersuchung von Webseiten. Die meisten Angebote konzentrieren sich auf die Sicherheit der verschlüsselten Verbindung zu einer Webseite, etwa \emph{Qualys}\footnote{\url{https://www.ssllabs.com/ssltest/}} und \emph{Mozilla}.\footnote{\url{https://observatory.mozilla.org/}} Darüber hinaus existieren Dienste, welche das Vorhandensein der relativ neuen HTTP"=Security"=Header überprüfen.\footnote{z.B. \url{https://securityheaders.io/}} Privatheitseigenschaften von Webseiten betrachtet  \emph{Webbkoll}.\footnote{\url{https://webbkoll.dataskydd.net/en/}}

Daneben gibt es Vorhaben, die eine festgelegte Liste von Webseiten untersuchen. So untersucht Helme regelmäßig die Verwendung der HTTP"=Security"=Header auf allen Webseiten der „Alexa Top 1 Million“\footnote{\url{https://scotthelme.co.uk/alexa-top-1-million-analysis-feb-2017}} und die schwedische Datenschutzorganisation \emph{dataskydd.net} sucht auf den Webseiten schwedischer Kommunen nach Privatheitsproblemen.\footnote{\url{https://dataskydd.net/kommuner-201704/}} %Eine Sammlung weiterer Datensätze findet sich bei \emph{\url{scans.io}}.
Im akademischen Umfeld gab es eine Reihe von Studien, die Aspekte der Sicherheit \cite{HolzAMKW16,MayerZSH16} und Privatheit \cite{Englehardt2016census} auf beliebten Webseiten untersuchen.
%Diese Studien bieten einen Überblick, differenzieren aber oft nicht zwischen verschiedenen Branchen von Webseiten (\cite{Englehardt2016census} ist hier eine Ausnahme).
Die Ergebnisse solcher einmalig durchgeführten Studien veralten jedoch recht schnell.
Insgesamt ist festzustellen, dass es bislang nicht ohne Weiteres möglich ist, aktuelle Benchmarks zu erstellen.
\section{PrivacyScore im Überblick}
\label{sec:overview}
\begin{figure}[t]
\centering
	\includegraphics[width=1\columnwidth]{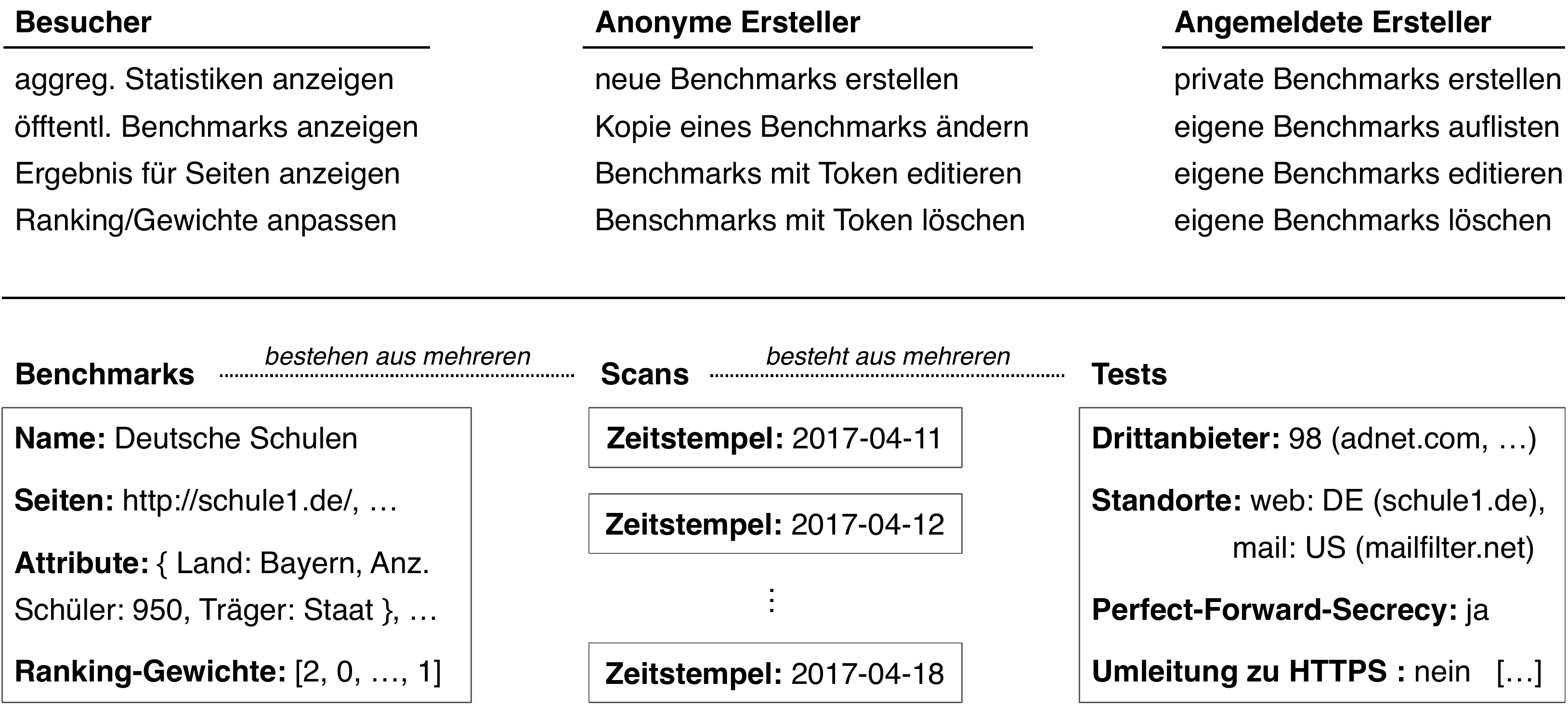} %
	\caption{\label{fig:usecases}Anwendungsfälle und Datenstrukturen}
\end{figure}

In diesem Abschnitt beschreiben wir die wichtigsten Funktionen des PrivacyScore"=Systems. Eine Übersicht über die wichtigsten Anwendungsfälle und Datenstrukturen gibt Abb.~\ref{fig:usecases}.
%
%
%\paragraph{Funktionalität}
PrivacyScore ist ein Webdienst, der im Auftrag seiner Nutzerinnen andere Webseiten analysiert. Wie bei anderen Diensten können einzelne Seiten analysiert werden. Bei PrivacyScore liegt der Fokus jedoch auf der Erstellung von \emph{Benchmarks}, in denen miteinander verwandte Webseiten verglichen werden. Ein Benchmark kann etwa die Seiten aller Firmen einer Branche enthalten oder (im Fall einer Datenschutz"=Aufsichtsbehörde) die Seiten aller Organisationen, die unter ihre Aufsicht fallen. Hinweise zur Datenbasis lassen sich in einer Freitext-Beschreibung hinterlegen.

%Die in einem Benchmark enthaltenen Seiten sind zur Sicherstellung der Konsistenz unveränderlich; ein Benchmark kann aber kopiert und als Basis für einen neuen Benchmark dienen.

Nachdem ein Benchmark erstellt wurde, wird er an die Scan"=Komponente übergeben, die die Seiten auf mehreren virtuellen Maschinen scannt. Die Ergebnisse werden auf der PrivacyScore"=Webseite veröffentlicht. Dort können Webseiten in einem Ranking miteinander verglichen und Detailergebnisse für einzelne Seiten abgerufen werden. Optional werden die Seiten jedes Benchmarks regelmäßig erneut gescannt.
Frühere Ergebnisse werden zur Dokumentation der zeitlichen Entwicklung aufbewahrt (\emph{Historie}).
%Durch Schlagworte (Tags) lassen sich ähnliche Benchmarks finden.

Desweiteren ist es möglich, die einzelnen Adressen mit bestimmten Attributen zu klassifizieren. Denkbar wäre zum Beispiel eine Unterscheidung zwischen den Webseiten der gesetzlichen und privaten Krankenkassen, die hier gegeneinander verglichen werden sollen.

PrivacyScore kann mit oder ohne Anmeldung verwendet werden. Angemeldete Nutzerinnen können ihre Benchmarks in ihrem Account speichern und haben leichteren Zugriff auf administrative Funktionen. Darüber hinaus können sie ihre Benchmarks als privat markieren.
%Nicht angemeldete Nutzerinnen erhalten beim Erstellen eines Benchmarks ein \emph{Token} zur späteren Verwaltung.

\paragraph{Flexibilität durch Nutzerorientierung}
PrivacyScore bietet zwei Funktionen, um sich an die Bedürfnisse der Nutzerinnen anzupassen: nutzerdefinierte Attribute und nutzerdefinierte Bewertungsschemata.
\emph{Nutzerdefinierte Attribute} erlauben es, die Liste von URLs mit frei definierbaren Attributen zu versehen, um verschiedene Klassen von Webseiten (z.\,B. gesetzliche und private Krankenkassen) voneinander zu unterscheiden und Zusammenhangsanalysen durchzuführen. Die Attribute werden beim Erstellen des Benchmarks festgelegt, können aber im Nachhinein modifiziert werden.
\emph{Nutzerdefinierte Bewertungsschemata} ermöglichen das Verändern der Bewertungskriterien für die Webseiten. Die Bewertung wird aus den Ergebnissen der verschiedenen Tests berechnet – da Nutzerinnen allerdings unterschiedliche Prioritäten haben können, welche Tests ihnen wichtig erscheinen, ist es wichtig, ihnen die Möglichkeit zu geben, die Gewichtung der Tests zu beeinflussen. Das verwendete Bewertungsschema kann jederzeit modifiziert werden, wobei entweder aus einer Liste vordefinierter Schemata ausgewählt oder ein neues Schema definiert werden kann.
Aus der gewichteten Bewertung können verschiedene vereinfachte Darstellungen (Schulnoten, Ampel, etc.) abgeleitet werden, die eine schnelle Bewertung der Ergebnisse erlauben. Datenschutz"=Aufsichtsbehörden können mit dieser Funktion vermutete Fälle von Nicht"=Konformität mit rechtlichen Vorgaben für eine weitere Untersuchung markieren.

\paragraph{Offene Daten und Vertraulichkeit}
PrivacyScore erzeugt Transparenz und Aufmerksamkeit, indem es auf Sicherheits- und Privatheitsprobleme hinweist. Um die Verbreitung der Ergebnisse zu erleichtern, werden sie nicht nur für Menschen lesbar aufbereitet, sondern auch über eine Schnittstelle in maschinenlesbarer Form veröffentlicht.

Um die Daten von als \emph{privat} markierten Benchmarks angemessen zu schützen, werden alle Tests von Servern unter unserer Kontrolle durchgeführt, d.\,h. die URLs der untersuchten Webseiten und die Scan"=Ergebnisse werden nicht an andere Dienstanbieter übermittelt.
Professionelle Anwender wie Datenschutz"=Aufsichtsbehörden haben jedoch u.\,U. noch strengere Anforderungen an Sicherheit und Vertraulichkeit. Diese Anforderungen können umgesetzt werden, indem sie eine eigenständige PrivacyScore"=Instanz in ihrer eigenen Infrastruktur betreiben. Um dies zu ermöglichen wird der Programmcode von PrivacyScore im Sommer unter einer freien Lizenz veröffentlicht. Dies kommt auch der Weiterentwicklung von PrivacyScore zu Gute, da es Dritten die Möglichkeit gibt, neue Tests beizusteuern.

\paragraph{Implementation}
\label{sec:implementation}
PrivacyScore befindet sich aktuell im \emph{Alpha}"=Stadium, ein Prototyp ist unter \url{https://privacyscore.org/} abrufbar. Für die Tests nutzen wir etablierte Tools wie \emph{OpenWPM} \cite{Englehardt2016census} und \emph{testssl.sh}.\footnote{\url{https://testssl.sh}} Weitere Details finden sich in \cite{Maass2017}.

\section{Sicherheits- und Privatheitstests}
\label{sec:checks}

Im Folgenden stellen wir einen Teil der Tests vor, die in \cite{Maass2017} beschrieben sind.
%Wir unterscheiden zwischen \emph{Sicherheitstests} und \emph{Privatheitstests}.
%
Unsichere Webseiten setzen ihre Besucherinnen Risiken aus, etwa dem Abhören durch bösartige WLAN"=Betreiber. Unsere \textbf{Sicherheitstests} prüfen, ob Seitenbetreiber anerkannte Methoden zur Absicherung ihrer Webseite und des Datenverkehrs  verwenden. 
%Daher ist die Sicherheit einer Webseite für die Privatheit ihrer Besucherinnen von Bedeutung.
PrivacyScore prüft u.\,a., ob \ac{TLS} korrekt eingesetzt wird, d.\,h. ob geeignete Protokollversionen, kryptographische Verfahren und HTTP"=Header wie \ac{HSTS} verwendet werden.
%PrivacyScore prüft u.\,a., ob \ac{TLS} eingesetzt wird und wie es um die Qualität der verwendeten Konfiguration bestellt ist. Dazu werden etwa die angebotenen Protokollversionen, die angebotenen kryptographischen Verfahren und das Vorhandensein von Sicherheitsmechanismen wie \ac{HSTS} geprüft.
Darüber hinaus werden die Mailserver überprüft, indem u.\,a. Verfügbarkeit und Güte der Transportverschlüsselung (STARTTLS) abgeklärt wird. %sowie die Verwendung von \ac{SPF} und \ac{DMARC}
%\cite{RFC7208,RFC7489})
Untersucht werden auch die Einträge im \ac{DNS}, insbesondere die Verwendung von DNSSEC, und die Aktualität der Server"=Software, sofern feststellbar.

\textbf{Privatheitstests} untersuchen hingegen, ob eine Webseite aktiv die Privatheit ihrer Besucherinnen verletzt oder gefährdet, etwa durch die Verwendung von Analyse- und Werbediensten.
%Dabei können sensiblen Informationen an Dritte übertragen werden, wo sie zum Profiling verwendet werden können.
PrivacyScore erkennt diese Dienste in einem zweistufigen Verfahren: Zunächst werden auf einer Webseite alle von externen Anbietern eingebetteten Ressourcen gesammelt. Diese werden in einem zweiten Schritt dann mit einer Liste bekannter Werbe- und Analysedienste abgeglichen. PrivacyScore wird außerdem eine Reihe von Tracking"=Technologien erkennen (insbesondere Cookies, Flash"=Cookies, und Browser"=Fingerprinting).

Viele Webseiten nutzen \acp{CDN}, um die Zuverlässigkeit und Geschwindigkeit ihrer Webseite zu optimieren. Diese Dienste können allerdings ebenfalls eine Gefahr für die Privatheit und Sicherheit darstellen.\footnote{\url{https://blog.cloudflare.com/incident-report-on-memory-leak-caused-by-cloudflare-parser-bug/}} PrivacyScore versucht daher die Verwendung von \acp{CDN} zu erkennen.
Eine weitere relevante Information ist der \emph{Ort}, an dem eine Webseite betrieben wird. Interessant ist hier vor allem die Einbindung nicht-europäischer Dienstanbieter, da ein Seitenbetreiber in diesem Fall möglicherweise zusätzliche Pflichten zu erfüllen hat.

\paragraph{Anschaulichkeit und Handlungsempfehlungen} 
Die identifizierten Probleme werden mit den daraus resultierenden Bedrohungen veranschaulicht. So ist etwa das Abhören verschlüsselter Kommunikation durch einen WLAN"=Betreiber möglich, wenn eine Seite kein \ac{HSTS} verwendet (\emph{SSL"=Stripping}"=Angriff).
PrivacyScore hilft den Seitenbetreibern aber auch, die Probleme zu beheben. Dazu werden wir Konfigurationshilfen für beliebte Webserver und lokale Analysesysteme wie \emph{Piwik}\footnote{\url{https://piwik.org}} auflisten.
% die Seitenbetreibern dabei helfen können, ihre Seiten sicherer und privatheitsfreundlicher zu gestalten.
Wo dies nicht möglich ist, liefern wir Nutzerinnen Hinweise zum Selbstdatenschutz, z.\,B. durch die Installation von Browser"=Add"=ons, die Werbenetzwerke und Tracker blockieren.

\section{Ethische Fragestellungen}
\label{sec:considerations}

Der automatische Abruf von Webseiten ist eine weitverbreitete Praxis (vgl. Suchmaschinen und z.\,B. \url{https://archive.org/}). Dennoch ist die Analyse von Seiten mit PrivacyScore aus ethischer Sicht nicht völlig unbedenklich, da sie über ein bloßes Abrufen einer Seite hinausgeht.
%In diesem Abschnitt erörtern wir die zentralen Fragestellungen.

Unser Ziel ist es, Seitenbetreiber und Nutzerinnen zu unterstützen, ohne jedoch ein (neues) Werkzeug für Angreifer zu entwickeln.
Einige unserer Tests können allerdings Informationen aufzeigen, die auch für Angreifer nützlich sein können. Wie viele andere Sicherheitswerkzeuge auch ist PrivacyScore demnach ein Dual-Use-Werkzeug.
Daher sind wir bei der Implementierung von Tests und der Darstellung der Ergebnisse bestrebt, Varianten zu wählen, die einen Missbrauch der Ergebnisse unwahrscheinlich machen.

Desweiteren belasten die TLS"=Tests die Infrastruktur des Anbieters, da sie eine hohe Anzahl an Verbindungen aufbauen.
Wir erachten dies für vertretbar, da das Anbieten einer Webseite die Absicht impliziert, sie zum Abruf zur Verfügung zu stellen.
Wir gehen davon aus, dass das öffentliche Interesse an den Ergebnissen i.\,d.\,R. die (meist vernachlässigbaren) Kosten für den Seitenbetreiber überwiegt.
Nichtsdestotrotz müssen wir sicherstellen, dass durch unsere Untersuchungen die Verfügbarkeit einer Seite nicht beeinträchtigt wird.
Um zu verhindern, dass PrivacyScore unabsichtlich oder absichtlich zu einem Seitenausfall führt, setzen wir ein Rate"=Limiting"=System ein: Ein erneuter Scan einer Webseite kann erst nach 30 Minuten erfolgen.

\section{Rechtliche Fragestellungen}
\label{sec:legal}
Automatisiertes Scannen von Webseiten im Internet wirft eine Reihe von Rechtsfragen in verschiedenen Rechtsbereichen auf. In diesem Abschnitt sollen exemplarisch die Rechtsfragen der Zulässigkeit einer Untersuchung, der Bewertung der Untersuchungsergebnisse und ihrer Veröffentlichung untersucht werden.

\paragraph{Zulässigkeit der automatisierten Untersuchung von Webseiten}
Das Auslesen fremder Webseiten ohne die ausdrückliche Zustimmung des Betreibers ist nicht grundsätzlich unzulässig. Zuerst stellt sich allerdings die Frage, wem die zu untersuchenden Daten „gehören“, denn wäre jemand Eigentümer dieser Daten, könnte er diese nutzen und über sie verfügen und andere gemäß §\,903 des Bürgerlichen Gesetzbuches (BGB) gerade von dieser Nutzung ausschließen. Die Begründung von Eigentum kann gem. §\,903 BGB nur an Sachen i.\,S.\,v. §\,90 BGB bestehen. Da es Daten bereits an der körperlichen Eigenschaft fehlt, kann es weder ein Eigentumsrecht an Daten noch ein aus Eigentum abgeleitetes Recht auf die ausschließliche Nutzung von Daten geben \cite{Zech2015}. Dennoch können Daten unter verschiedenen gesetzlichen Voraussetzungen vor dem Zugriff und der Verwertung durch Dritte geschützt sein. Daten unterliegen einer Kommunikationsordnung, welche sich aus Regelungen in diversen Rechtsgebieten zusammensetzt und Umgangsrechte, Verfügungsberechtigungen und -beschränkungen hinsichtlich der Daten gewährt, ohne dabei eine Güterzuweisung vorzunehmen \cite{Rossnagel2014,Rossnagel2017}.
Relevant werden können, soweit Gegenstand personenbezogene Daten von natürlichen Personen sind, Regelungen aus dem Datenschutzrecht, auch Regelungen aus dem Urheberrecht, sofern es sich um persönliche geistige Schöpfungen handelt, weiterhin auch einzelne Regelungen aus dem Strafgesetzbuch (StGB) oder dem Gesetz gegen den unlauteren Wettbewerb (UWG). Darüber hinaus ist auch ein deliktischer Abwehranspruch gegen Störungen aus dem Recht am eingerichteten und ausgeübten Gewerbebetrieb i.\,S.\,v. §\,823~Abs.\,1 BGB denkbar.

Einige Webseitenbetreiber verbieten in ihren Nutzungsbedingungen das automatisierte Durchsuchen und Auswerten der von ihnen zur Verfügung gestellten Daten. Während manche deutsche Gerichte ein hierzu berechtigendes „virtuelles Hausrecht“ generell bejahen\footnote{LG Hamburg v. 28.08.2008, Az.\,315~O~326/08.} oder dies von der Art der Webseitenausgestaltung abhängig machen,\footnote{OLG Köln v. 25.08.2000, Az.\,19~U~2/00; LG Ulm v. 13.01.15, Az.\,2~O~8/15; OLG Hamm v. 10.06.2008, Az.\,4~U~37/08.} spricht gegen die Übertragung der Befugnis zur Ausübung des Hausrechts von Räumen und Grundstücken auf Webseiten die nicht vergleichbare Interessenlage, da ursprünglich gerade absolute Rechtspositionen wie Besitz und Eigentum Gegenstand des Schutzes sein sollten und der Zweck einer Webseite gerade darin besteht, sie potentiellen Nutzerinnen zu öffnen und von diesen zur Kenntnis genommen zu werden.\footnote{OLG Frankfurt v. 05.03.2009, Az.\,6~U~221/09; OLG Hamburg v. 24.10.2012, Az.\,5~U~38/10.} Dem Webseitenbetreiber steht es dennoch frei, die Befugnisse über die von ihm zur Verfügung gestellten Daten im Rahmen vertraglicher Vereinbarungen zu regeln. Rechtlich verbindlich wird die jeweilige Vorgabe in den Nutzungsbedingungen aber nur dann, wenn zuvor eine Registrierung und damit ein Vertragsschluss unter ausdrücklicher Anerkennung der Nutzungsbedingungen erfolgt.\footnote{LG München v. 25.10.2006, Az.\,30~O~11973/05; OLG Frankfurt v. 05.03.2009, Az.\,6~U~221/09.} Ist der Zugang dagegen auch ohne Anmeldung möglich, kommt den jeweiligen Nutzungsbedingungen ebenso wie allen einseitigen Erklärungen über gewollte Nutzungsbeschränkungen keine verbindliche Rechtswirkung für nicht registrierte Besucherinnen der Webseite zu.\footnote{OLG Frankfurt v. 05.03.2009, Az.\,6~U~221/09; OLG Hamburg v. 24.10.2012, Az.\,5~U~38/10.} Eine solche technisch"=automatisierte Registrierung unter Anerkennung der gestellten Nutzungsbedingungen durch die PrivacyScore"=Technik ist nicht vorgesehen. 

Ein Großteil der PrivacyScore"=Tests basiert lediglich auf Metadaten, die beim Abruf einer Seite entstehen. Einige Tests profitieren jedoch von einer Analyse des Quelltexts einer Seite. Ausnahmsweise könnten ausschließliche Nutzungsrechte am Quelltext dadurch entstehen, dass diesem urheberrechtlicher Schutz zukommen würde. In diesem Fall müssten betroffene Webseitenbetreiber die Datenverwendung nicht dulden und hätten entsprechende Unterlassungs- und Schadensersatzansprüche.
%Grundsätzlich kann unter bestimmten Bedingungen einzelnen Webseiten Urheberrechtsschutz zukommen.
Dies ist davon abhängig, ob die Webseite als Werk i.\,S.\,v. §\,2~Abs.\,1 UrhG oder als eine geschützte Datenbank i.\,S.\,v. §\,4~Abs.\,2 UrhG qualifiziert werden kann. Hierfür ist notwendig, dass die Gestaltung der Webseite die „geistige Schöpfungshöhe“ i.\,S.\,v. §\,2~Abs.\,2 UrhG erreicht. Dies kann nur im Einzelfall beurteilt werden. Regelmäßig ist der Quelltext der Webseite jedoch nicht urheberrechtlich geschützt, da die Rechtsprechung die nötige besondere „schöpferische Höhe“ i.\,S.\,v. §\,2~Abs.\,2 UrhG für Webseiten nur in wenigen Ausnahmefällen anerkennt.\footnote{OLG Hamburg v. 29.02.2012, Az.\,5~U~10/10; OLG Celle v. 08.03.2012, Az.\,13~W~17/12; OLG Rostock v. 27.06.2007, Az.\,2~E~12/07.} Auch, wenn die Datenbank als geistiger Schöpfungsakt nicht urheberrechtlich geschützt ist, kann für den Hersteller der Datenbank nach §\,87b~Abs.\,1~Satz~1 UrhG ein Leistungsschutzrecht bestehen. Ginge man davon aus, dass die im Quelltext enthaltenen Informationen eine Sammlung von Daten i.\,S.\,v. §\,87a UrhG enthalten, käme es für die Annahme eines Leistungsschutzrechts weiterhin darauf an, ob beim Betreiber eine wesentliche Investition zur Erstellung erforderlich war. Allerdings räumt §\,87b~Abs.\,1~Satz~1 UrhG ein Recht auf Untersagung der Nutzung nur ein, soweit die Nutzerin wesentliche Teile der Datenbank durchsucht. Soweit die Webseite, wie bei PrivacyScore, lediglich auf deren technische Umsetzung untersucht wird, stellen diese Informationen keine wesentlichen Teile der Datenbank dar. Für eine nach §\,87b~Abs.\,1~Satz~2 UrhG unzulässige Vervielfältigung, die auch für unwesentliche Teile gilt, fehlt es an einer wiederholten und systematischen Handlung, da sich die Nutzung der Daten durch PrivacyScore im Rahmen einer normalen Auswertung hält. Quelltexte von Webseiten genießen regelmäßig auch keinen Sonderrechtsschutz als Computerprogramme i.\,S.\,v. §\,2~Abs.\,1~Nr.\,1, 69a~Abs.\,1 UrhG. Dies würde eine Folge von Befehlen, denen Auswirkungen auf den Programmablauf zukommen, voraussetzen. Den HTML"=Befehlen im Quelltext komme dagegen nur eine beschreibende Funktion zu, um die Darstellung der Inhalte im Browser zu ermöglichen.\footnote{OLG Rostock v. 27.06.2007, Az.\,2~W~12/07; OLG Frankfurt v. 22.03.2005, Az.\,11~U~64/04.}

Die Untersuchung durch PrivacyScore ist auch wettbewerbsrechtlich unbedenklich, solange es infolge seines Einsatzes nicht zu einer tatsächlichen Störung des Betriebsablaufs der Webseite, etwa einer Beeinträchtigung der Verfügbarkeit, kommt. Andernfalls könnte unter der Voraussetzung der Mitbewerbereigenschaft i.\,S.\,v. §\,2~Nr.\,3 UWG eine unlautere Behinderung i.\,S.\,v. §\,4~Nr.\,4 UWG vorliegen. Wie bereits in Abschnitt~\ref{sec:considerations} dargestellt, wird durch ein Rate"=Limiting"=System vermieden, dass die Untersuchung durch PrivacyScore negative Auswirkungen auf die Verfügbarkeit des Servers durch technische Überlastung für andere Nutzerinnen haben kann. Eine Störung der Funktionsfähigkeit i.\,S.\,v. §\,4~Nr.\,4 UWG soll damit ausgeschlossen werden. Mit der gleichen Begründung kann auch ein deliktischer Anspruch, gestützt auf eine rechtswidrige Verletzung des Rechts am eingerichteten und ausgeübten Gewerbebetrieb i.\,S.\,v. §\,823~I BGB, mangels eines betriebsbezogenen Eingriffs, der über eine bloße Belästigung hinausgeht, vermieden werden.

Sofern PrivacyScore zur Untersuchung der Webseite nicht eine vom Berechtigten errichtete „besondere Zugangssicherung“ überwindet, bestehen auch in strafrechtlicher Hinsicht keine Bedenken gegen dessen Einsatz. In Betracht käme allenfalls der Tatbestand des §\,202a StGB, welcher Daten schützt, „die gegen unberechtigten Zugang besonders gesichert sind“, indem die Zugangsverschaffung gerade „unter Überwindung der Zugangssicherung“ unter Strafe gestellt wird. Mithilfe von PrivacyScore wird nur der Quelltext der Webseite untersucht, der für jedermann öffentlich zugänglich ist ohne dabei eine Zugangssicherung zu überwinden. 

Zusammenfassend ist festzustellen, dass soweit kein Sonderrechtsschutz besteht und die Webseite rechtlich und technisch frei zugänglich ist, d.\,h. die Nutzung der zu untersuchenden Daten nicht von der vorherigen Akzeptanz der Nutzungsbedingungen abhängig gemacht und vom Webseitenbetreiber keine technischen Schutzmaßnahmen installiert wurden,\footnote{BGH v. 22.06.2011, Az.\,I~ZR~159/10; BGH v. 17.03.2003, Az.\,I~ZR~259/00.} das automatisierte Untersuchen der Webseiten durch PrivacyScore zulässig ist.

\paragraph{Rechtliche Bewertung der Untersuchungsergebnisse}

Die bloße Feststellung des technischen Untersuchungsergebnisses allein ist zur Bewertung der rechtskonformen Gestaltung der Webseite wenig aussagekräftig. Zur tatsächlichen Beurteilung der Rechtskonformität ist immer eine weitergehende Betrachtung des Einzelfalls erforderlich. Dies soll im Folgenden am Beispiel von eingebundenen Analysediensten und fehlender bzw. veralteter Verschlüsselungssoftware dargestellt werden.

Das Einbinden von Analysediensten ist nicht per se nur nachteilhaft für Nutzerinnen, es besteht aber durchaus die Gefahr des Missbrauchs. Die Rechtmäßigkeit von Analysediensten, wie Google Analytics, auf Webseiten wird seit jeher von den Datenschützern kritisch betrachtet, da sie befürchten, dass mittels dieser Analyse umfangreiche Profile über identifizierbare Nutzerinnen angelegt werden könnten. Neben der Zuordnung und Speicherung von Nutzungsdaten (wie die zuletzt besuchte Seite und der verwendete Browser), die zunächst rechtlich unbedenklich sind, ermöglichen zentrale Analysedienste wie Google Analytics die Möglichkeit, die Nutzungsaktivitäten über die ebenfalls abgespeicherte IP"=Adresse zusammen zu führen und webseitenübergreifend einer bestimmten Nutzerin zuzuordnen. Eine datenschutzkonforme Nutzung von Analysediensten wie Google Analytics ist daher nur zulässig, wenn eine Anonymisierung der IP"=Adresse – etwa in Form einer Kürzung – gewährleistet wird, da aufgrund dieser Kürzung nach überwiegender Ansicht der Datenschützer\footnote{Beschluss des Düsseldorfer Kreises vom 26./27.11.2009 über die „datenschutzkonforme Ausgestaltung von Analyseverfahren“.} der direkte Personenbezug entfalle. Soweit nach der jüngsten Rechtsprechung\footnote{BGH v. 16.05.2017, Az.\,VI~ZR~135/13; EuGH v. 19.10.2016, Az.\,C-582/14.} dynamische IP"=Adressen personenbezogene Daten darstellen, dürfen sie nur verarbeitet werden, wenn dies entweder gesetzlich gestattet ist oder die Betroffene eingewilligt hat.

Die gesetzliche Grundlage für den Umgang mit Nutzungsdaten ist derzeit noch das nationale Telemediengesetz (TMG), speziell §\,15~Abs.\,3 TMG, der auch im Falle einer verkürzten IP anwendbar ist.\footnote{LG Frankfurt v. 18.02.2014, Az.\,3-10~O~86/12.}
Nach dieser Vorschrift ist die Erstellung von pseudonymisierten Nutzungsprofilen zu festgelegten Zwecken auch ohne die Einwilligung der betroffenen Nutzerin zulässig, soweit der Nutzerin gem. §\,15~Abs.\,3~Satz~2 TMG ein Recht zum Widerspruch („Opt"=out“) gegen diese Erstellung eingeräumt und sie auf die Möglichkeit des Widerspruchs auch hingewiesen wird.\footnote{BGH v. 16.07.2008, Az.\,VIII~ZR~348/06.} Um beurteilen zu können, ob der Webseitenbetreiber seinen Hinweispflichten ausreichend nachgekommen ist, wäre eine Betrachtung des Impressums und der Datenschutzerklärung notwendig. 

Etwas anderes kann sich möglicherweise zukünftig aufgrund des Entwurfs der Verordnung über Privatsphäre und elektronische Kommunikation (e-Privacy-VO-E) ergeben. Nach Art.\,8 Abs.\,1 des Verordnungsentwurfs, ist „jede vom betreffenden Endnutzer nicht selbst vorgenommene Nutzung der Verarbeitungs- und Speicherfunktionen“ grundsätzlich verboten, es sei denn, der Endnutzer hat seine Einwilligung gegeben oder es greifen die übrigen Ausnahmen von Abs.\,1 lit.\,a–d. Ob unter die Ausnahme nach Art.\,8 Abs.\,1 lit.\,d auch externe Analysedienste fallen, hängt von der Auslegung der Einschränkung, „sofern der Betreiber (...) diese Messung durchführt“, ab. Dies könnte problematisch sein, da externe Dienste gerade nicht identisch mit dem Webseitenbetreiber sind, der den vom Endnutzer gewünschten Dienst anbietet. Die Folge wäre, dass zukünftig das Nutzen von Analysediensten generell von der vorherigen Einwilligung („Opt"=in“) der betroffenen Nutzerin abhängig wäre. Zudem hätten Nutzerinnen zukünftig gem. Art.\,9 Abs.\,3 des Verordnungsentwurfs ein jederzeitiges Widerrufsrecht bezüglich der einmal erteilten Einwilligung, an das sie halbjährlich erinnern werden müssten. Im Übrigen würden sich die Anforderungen der Einwilligung gem. Art.\,9 Abs.\,1 des Verordnungsentwurfs nach Art.\,7 und 4~Nr.\,11 DSGVO richten.

Ähnlich stellt es sich bei fehlender Verschlüsselungssoftware dar, da eine generelle Verschlüsselungspflicht für Webseitenbetreiber nicht ausdrücklich normiert ist. Vorschriften zur Datensicherheit finden sich derzeit u.\,a. in §\,9 Bundesdatenschutz (BDSG) und dessen Anlage sowie im für Telemedien spezielleren §\,13~Abs.\,7 TMG. Da die Datenschutz"=Grundverordnung (DSGVO) und die e-Privacy-VO als europäische Verordnungen gem. §\,288~Abs.\,2 AEUV mit Geltungsbeginn am 25.5.2018 unmittelbar in den Mitgliedsstaaten anwendbar sein werden, werden sie diese nationalen Regelungen im Rahmen des Anwendungsvorrangs verdrängen. Aus Erwägungsgrund 5 des Entwurfs der e-Privacy-VO geht hervor, dass diese gegenüber der DSGVO spezieller sein soll, wenn ihr Anwendungsbereich betroffen ist. Gleichzeitig verweist Erwägungsgrund 37 des Verordnungsentwurfs hinsichtlich der Bewertung der Datensicherheit auf Art.\,32 DSGVO, in dem i.\,V.\,m. Art.\,5~Abs.\,1~lit.\,f DSGVO die Datensicherheit als allgemeiner Grundsatz normiert ist. Art.\,32 DSGVO verpflichtet den Verantwortlichen (Art.\,4~Nr.\,7 DSGVO) sowie den Auftragsverarbeiter (Art.\,4~Nr.\,8 DSGVO) „geeignete technische und organisatorischen Maßnahmen“ zur Herstellung eines „angemessenen Datenschutzniveaus“ zu gewährleisten. Neben der Pseudonymisierung nennt der Normgeber in Art.\,32~Abs.\,1 lit.\,a DSGVO auch die Verschlüsselung als eine der bevorzugten Maßnahmen. Ebenso wie die Pseudonymisierung, ist die Verschlüsselung jedoch nicht in jedem Einzelfall zwingend geboten. Welche Maßnahmen sich als erforderlich und angemessen erweisen, um ein „dem Risiko angemessenes Schutzniveau zu gewährleisten“, muss vielmehr im Rahmen einer Gesamtabwägung an das Einzelfallrisiko angepasst werden. Hierfür ist im Rahmen des Art.\,32~Abs.\,1 DSGVO neben der Schwere und Eintrittswahrscheinlichkeit des Risikos für die Rechte und Freiheiten der von der Verarbeitung Betroffenen, das sich insbesondere an der Schutzbedürftigkeit der einzelnen gespeicherten personenbezogenen Daten orientiert, auf der anderen Seite sowohl das technisch Machbare nach dem „Stand der Technik“ als auch die wirtschaftliche Belastung des Verarbeiters, insbesondere die Implementierungskosten, mit einzubeziehen. Lediglich das wirtschaftlich Zumutbare wird erwartet. Dies kann im Einzelfall auch dazu führen, dass ein Verantwortlicher keine Maßnahmen ergreifen muss.

\paragraph{Rechtliche Herausforderungen hinsichtlich der Veröffentlichung der Ergebnisse}

Auf der PrivacyScore"=Webseite können sowohl Detail"=Ergebnisse für einzelne Webseiten als auch vergleichende Rankings eingesehen werden. Bei den Detail"=Ergebnissen handelt es sich um Tatsachenbehauptungen, die überprüfbar und damit dem Beweis zugänglich sind. Die erstellten Rankings dürften hingegen auf diesen Tatsachen beruhende Werturteile darstellen. Beide sind nach ständiger Rechtsprechung gleichermaßen von der Meinungsfreiheit aus Art.\,5~Abs.\,1 GG umfasst, soweit es sich um wahre Tatsachen handelt.\footnote{BVerfG v. 09.10.1991, Az.\,1~BvR~1555/88; BVerfG v. 13.04.1994, Az.\,1~BvR~23/94.} Die zugehörigen Scan"=Rohdaten werden dazu (soweit zulässig) archiviert.

Da mit der Veröffentlichung nachteiliger Bewertungen auch negative Auswirkungen für den Webseitenbetreiber, wie Imageverluste, verbunden sein können, stellt sich die Frage, unter welchen Voraussetzungen sich Betreiber gegen die Veröffentlichung wehren können. 

Veröffentlichungen über Tatsachen und Werturteile sind solange zulässig und von den Bewerteten hinzunehmen, wie sie keinen rechtswidrigen Eingriff in das Persönlichkeitsrecht des Betreibers darstellen. Ein solcher ist anzunehmen, wenn im Rahmen einer Abwägung das Schutzinteresse des von der Äußerung Betroffenen auf Schutz seiner Persönlichkeit aus Art.\,1~Abs.\,1 GG i.\,V.\,m. Art.\,2~Abs.\,1 GG, die schutzwürdigen Belange an der Veröffentlichung aus Art.\,5~Abs.\,1 GG überwiegen. Für inländische Unternehmen kann sich ein Persönlichkeitsrecht aus Art.\,2~Abs.\,1 GG i.\,V.\,m. Art.\,19~Abs.\,3 GG und Art.\,12~Abs.\,1 GG ergeben. Dies gilt jedoch nur in abgeschwächter Form, da Unternehmen und Unternehmer, die sich bewusst nach außen hin darstellen, kritische Bewertungen eher hinnehmen müssen.\footnote{BGH v. 29.01.2002, Az.\,VI~ZR~20/01.} Juristische Personen des öffentlichen Rechts sind grundsätzlich nicht Träger von Persönlichkeitsrechten, ein zivilrechtlicher Schutz steht ihnen daher nur mittelbar zu, sofern sie Adressat strafbarer Äußerungen i.\,S.\,v. §§\,185\,ff. StGB werden und die konkrete Aussage geeignet ist, die Behörde in ihrer Funktion zu beeinträchtigen. 

Eine Beurteilung eines rechtswidrigen Eingriffs unter Abwägung der Interessen ist grundsätzlich nur im Einzelfall möglich. Die Veröffentlichung einer kritischen Bewertung der Webseite kann für den Betroffenen zu einer Rufbeeinträchtigung führen und ihn nicht unerheblich belasten. Da die untersuchten Webseiten aber öffentlich sind, werden diese regelmäßig der Sozialsphäre des Betroffenen zuzuordnen sein, d.\,h. beruflichen, politischen oder ähnlichen Tätigkeiten, in denen Menschen im sozialen Austausch stehen. Beeinträchtigungen in diesem Bereich unterliegen zumeist einem schwachen Schutz. Im Rahmen der Interessenabwägung bei betroffenen Unternehmern betont die Rechtsprechung\footnote{BGH v. 23.09.2014, Az.\,VI~ZR~358/13; BGH v. 23.06.2009, Az.\,VI~ZR~196/08; OLG Hamburg v. 18.01.2012, Az.\,5~U~51/11.} regelmäßig das Informationsinteresse der Allgemeinheit. Insbesondere vor dem Hintergrund der freien Wahl, die untersuchten Webseiten je nach Ergebnis zu nutzen oder dies zu vermeiden, ist das Interesse der Öffentlichkeit an den Informationen über privatrechtliche Risiken beim Besuch der Webseiten nicht unerheblich. Ein Recht des Betroffenen wird daher regelmäßig nicht das Kommunikationsinteresse überwiegen. Bei Privatpersonen wird im Rahmen der Interessenabwägung häufig zusätzlich auf die Breitenwirkung der Äußerung abgestellt: je größer der Adressatenkreis, desto belastender die Äußerung.\footnote{BGH v. 23.06.2009, Az.\,VI~ZR196/08.} Eine Risikoreduzierung für den Betroffenen könnte in diesen Fällen dadurch erreicht werden, dass die Ergebnisse nur einem beschränkten Adressatenkreis zur Verfügung zu stellen.

Aus Art.\,17~Abs.\,1 DSGVO könnte dem Betroffenen in diesem Zusammenhang ein Löschungsanspruch zustehen, soweit die gespeicherten Daten personenbezogen sind und deren Speicherung unzulässig ist. Eine Datenverarbeitung ist gem. Art.\,5~Abs.\,1~lit.\,a und Art.\,6~Abs.\,1 DSGVO dann rechtmäßig, wenn eine Einwilligung des Betroffenen vorliegt, was regelmäßig nicht der Fall sein wird, oder eine andere Bedingung des Art.\,6~Abs.\,1 DSGVO erfüllt ist. In Betracht käme als Erlaubnistatbestand  Art.\,6~Abs.\,1~UAbs.\,1~lit.\,f DSGVO. Dazu müsste die Verarbeitung der Daten „zur Wahrung der berechtigten Interessen des Verantwortlichen oder eines Dritten erforderlich“ sein, „sofern nicht die Interessen oder Grundrechte der betroffenen Person überwiegen“. Im Rahmen der durchzuführenden Interessenabwägung kann aufgrund der eben angeführten Argumente regelmäßig ein Überwiegen des Kommunikationsinteresses gegenüber dem Recht des Betroffenen auf informationelle Selbstbestimmung angenommen und ein Löschungsanspruch verneint werden.

Sofern sich die negative Bewertung im Rahmen von Art. 5 Abs. 1 GG bewegt, scheiden auch wettbewerbsrechtliche Ansprüche gem. §\,4~Nr.\,1 und 2 i.\,V.\,m. §\,8 UWG aus.

\section{Schlussbemerkungen}
\label{sec:conclusion}
Der sichere und datenschutzfreundliche Betrieb einer Webseite ist aufwändig, 
%Einerseits müssen Betreiber permanent ihre Sicherheitsvorkehrungen auf dem aktuellen Stand halten, andererseits müssen sie dabei auch die Privatsphäre ihrer Besucherinnen respektieren.
meist mit zusätzlichen Kosten verbunden oder nicht mit dem Geschäftsmodell eines Seitenbetreibers vereinbar. Daher ist das Sicherheits- und Privatheitsniveau vieler Webseiten derzeit unzureichend.
Das Ziel des PrivacyScore"=Projekts besteht darin, Transparenz herzustellen und Öffentlichkeit zu generieren.
PrivacyScore erstellt dazu automatisierte Vergleiche der Sicherheits- und Privatheitseigenschaften mehrerer Webseiten und zeigt die Ergebnisse in Form von regelmäßig aktualisierten Ranglisten an.
Sowohl die zu überprüfenden Webseiten als auch die Bewertungskriterien können dabei von den Nutzerinnen festgelegt werden.

Juristisch ist die Verwendung von PrivacyScore in seiner aktuell geplanten Form zulässig, soweit die untersuchte Webseite frei zugänglich ist und kein Sonderrechtsschutz besteht. Die Veröffentlichung der Ergebnisse ist vom Schutz der Meinungsfreiheit umfasst, sofern diese keine unwahren oder beleidigenden Äußerungen enthalten. Regelmäßig ist auch keine Einwilligung des Betreibers der untersuchten Webseite erforderlich.

%Zusammenfassend kann festgestellt werden, dass allein der Umstand, dass sich PrivacyScore über den von den Webseitenbetreibern geäußerten Willen gegen eine automatisierte Durchsuchung hinwegsetzt, für eine Unzulässigkeit nicht ausreicht. Soweit nicht ausnahmsweise Sonderrechtsschutz besteht und die Webseite rechtlich und technisch frei zugänglich ist, d. h. die Nutzung der zu untersuchenden Daten nicht von der vorherigen Akzeptanz der Nutzungsbedingungen abhängig gemacht und vom Webseitenbetreiber keine technischen Schutzmaßnahmen installiert wurden,\footnote{BGH v. 22.06.2011, Az.\,I~ZR~159/10; BGH v. 17.03.2003, Az.\,I~ZR~259/00.} ist das automatisierte Untersuchen der Webseiten durch PrivacyScore zulässig.

%Die Veröffentlichung der Untersuchungsergebnisse/Bewertungen ist grundsätzlich vom Schutz der Meinungsfreiheit umfasst. Etwas anderes gilt nur, würden die Ergebnisse unwahre Tatsachen oder beleidigende Äußerungen enthalten. Auch unter dem Aspekt des Datenschutzes und des Rechts auf informationelle Selbstbestimmung ist die Veröffentlichung ohne Einwilligung rechtmäßig, soweit sich die Bewertungen auf Tätigkeiten aus dem Bereich der Sozialsphäre beschränken. Sofern sich die negative Bewertung im Rahmen von Art.\,5~Abs.\,1 GG bewegt, scheiden auch wettbewerbsrechtliche Ansprüche gem. §\,4~Nr.\,1~und~2 i.V.m. §\,8~UWG aus.

PrivacyScore ist Open"=Source"=Software. Das System kann daher auch innerbetrieblich eingesetzt werden und beispielsweise die Arbeit von Datenschutz"=Aufsichtsbehörden unterstützen. Die auf der öffentlich angebotenen PrivacyScore"=Seite  erzeugten Datensätze werden darüber hinaus als Rohdaten für Forschungszwecke zur Verfügung gestellt.

\paragraph{Danksagung}
\label{sec:Acknowledgments}

Teile dieser Arbeit wurden in Forschungsteilbereichen C.1 und C.2 innerhalb des GRK 2050 „Privacy and Trust for Mobile Users“ durch die DFG finanziert. Die Autoren danken Marvin Hebisch, Nico Vitt und Pascal Wichmann für die Implementation des Prototypen, sowie den Teilnehmenden des Workshops \emph{PET-CON 2017.1} und Mitgliedern von Digitalcourage~e.\,V. für ihre Ideen.

%%% Angabe der .bib-Datei (ohne Endung) / State .bib file (for BibTeX usage)
\bibliography{bibliography} %\printbibliography if you use biblatex/Biber
\end{document}